\documentstyle[prd,aps]{revtex}
\begin{document}

\title{
Is there a connection between no-hair behavior and universality in
gravitational collapse?}

\author{Jorge Pullin}
\address{
Center for Gravitational Physics and Geometry\\
Department of Physics, 104 Davey Lab\\
The Pennsylvania State University\\
University Park, PA 16802}
\maketitle
\begin{abstract}

We apply linear perturbation theory to the study of the universality and
criticality first observed by Choptuik in gravitational collapse.  Since
these are essentially nonlinear phenomena our attempt is only  a rough
approximation.  In spite of this, universal behavior of the final black
hole mass is observed with an exponent of 1/2, slightly higher than the
observed value of 0.367.  The universal behavior is rooted in the
universal form that in-falling perturbations on black holes have at
the horizon.

\end{abstract}
\vspace{-7.8cm}
\begin{flushright}
\baselineskip=15pt
CGPG-94/9-2  \\
gr-qc/9409044\\
\end{flushright}
\vspace{7cm}

If one considers the collapse of a spherically symmetric scalar field in
general relativity, two possible end-results are expected.  Either the
initial data manages in the time evolution to create a big enough mass
concentration such that a black hole forms or whatever scalar field was
present is dispersed to infinity, leaving a flat spacetime as a result.
If one considers one-parameter families of initial data, there will be
in the parameter range two distinct regions, one corresponding to the
formation of a black hole and another to dispersion to infinity.
Christodoulou and others \cite{Chr,GoPi} had raised the question of what
was the behavior of the final black hole mass as a function of one such
parameter.  For values of the parameter below a "critical" value for
which black holes start forming, the final mass is zero, since no black
hole forms.  When the critical value is reached, is there a ``jump" in
the mass, i.e., is there a minimal mass for which one can form a black
hole? This was highly unlikely, since the problem has no scale.
Choptuik \cite{Ch} has confirmed this through numerical simulations.  He
finds no ``mass gap", but actually finds the final black hole mass is a
{\em universal} function of the parameter value of the initial data,
\begin{equation}
M_{BH} = k (p -p_*)^\beta
\end{equation}
where $k$ is a constant that may be different for different one-parameter
families of data, $p$ is the parameter value chosen for the initial data
and $p_*$ is the critical parameter value at which black holes start to
form. The exponent is {\em universal} (i.e. family-independent) and was
experimentally determined by Choptuik to be $\beta=0.367$. This value
was later confirmed with a different (less sophisticated)
kind of code \cite{gpp}.

In a separate set of experiments, Abrahams and Evans \cite{AbEv}
considered the
implosion of an axially symmetric gravitational wave (in vacuum, no scalar
field present). They find a similar behavior with the same exponent.
Finally, Coleman and Evans \cite{CoEv}
considered the collapse of a radiative fluid
$p=\rho/3$ and also found a similar behavior with the same exponent.

Little theoretical progress has been made towards explaining this
universal behavior and in particular the value of the exponent. Several
authors \cite{Br,Na,Hu}
have noticed that there exist in the literature exact self-similar
solutions for the collapse of scalar fields. These solutions are not
quite geared towards describing the effects observed by Choptuik, since
they are not asymptotically flat. In spite of this they yield a critical
behavior with an exponent of 1/2. An exponent of 1/2 has been also observed
in a rather unrelated context, the collapse of $1+1$ dimensional
black holes with quantum corrections by Strominger and Thorlacius
\cite{StTh}.

The purpose of this note is to explore the application of perturbation
theory to these phenomena.  Our approach is crude, and the results can
only be taken as suggestive.  It is remarkable however, that we find a
universal exponent that is close to the observed value (1/2 instead of
0.367), and that the root for the universality is the same as that of
the no-hair behavior of black holes: the universal form that fields
falling into the horizon take (quasinormal ringing).

Suppose one prepares a set of initial data for the Choptuik experiment,
with a certain value of the parameter $p_1$ (although it is not quite needed,
the reader can imagine a Gaussian of amplitude $p_1$ for concreteness).
Let us also assume $p_1>p_*$ and therefore, according to Choptuik,
a black hole forms, with mass,
\begin{equation}
M_1 = k (p_1-p_*)^\beta.
\end{equation}

Suppose now a second set of data is prepared, of parameter value $p_2>p_1$,
with $p_2$ differing little from $p_1$, i.e., $p_2-p_1<p_1-p_*$. Again, we
form a black hole of mass,
\begin{equation}
M_2 = k (p_2-p_*)^\beta
\end{equation}
and we are assuming the two sets of data are in the same family, so the
constant $k$ is the same.  We now would like to view the second set of
data as a ``perturbation" of the first set.  Assuming their parameter
separation to be small, i.e. $\Delta p =p_2-p_1 << p_1-p_*$,
we can write for the final mass difference
$\Delta M =M_2-M_1$,
\begin{equation}
\Delta M = k \beta (p_2 -p_1) (p_1 - p_*)^{\beta-1}
\end{equation}
which can be written as,
\begin{equation}
\Delta M = k^{1\over \beta} \beta \Delta p M_1^{\beta-1\over \beta}.
\end{equation}

Therefore, if there would be a way of estimating the excess mass that
falls into a black hole due to an increase of the parameter value of the
initial data, one could estimate the exponent $\beta$. Alternatively,
if one could estimate $\Delta M$ as a function of the black hole mass
$M_1$ one could also get  a value for the parameter $\beta$.

We will now provide such an estimate using perturbation theory.  We
start by making a rough assumption: the spacetime geometry for the
collapse of the first initial data set can be approximated by the
spacetime geometry of a static, ever-existing, Schwarzschild black
hole of mass $M_1$.  This is an uncontrolled and vague approximation.
One expects it in general grounds to be good if $p_1$ is far away from
criticality, so one forms a black hole with most of the initial data
mass and there is little outgoing radiation. Also one should realize
that approximating a time-dependent spacetime by a static one can only
work in the region where the time-dependent spacetime is not rapidly
changing.  In fact, numerical studies of perturbations on dynamical
backgrounds have shown that the approximation can work remarkably well
\cite{gpp}. However, there is no real justification to apply this
approximation in the case we are interested since the attention should
be focused near criticality, when the background differs considerably
{}from an ever-existing black hole.  What follows therefore should only
be taken as suggestive and speculative.

If one assumes the spacetime determined by the initial data set with
parameter $p_1$ is a Schwarzschild black hole of mass $M_1$, estimating
$\Delta M$ becomes a problem in perturbations of black holes: given a
perturbation incident on a black hole, how much mass falls into the hole
and how much mass is radiated away? To estimate this we note that
perturbations on black hole backgrounds are determined by the Zerilli
function $\Psi$ (to summarize, one takes $g_{\mu\nu} = {\rm
Schwarzschild}_{\mu\nu} +h_{\mu\nu} $ and finds that all the information
in $h_{\mu\nu}$ can be parameterized by a function $\Psi$, see \cite{CPM}
for details).  This function satisfies a Klein-Gordon-like equation coupled
to a potential called the Zerilli equation.  The ``radial" variable in
this equation is a ``tortoise" coordinate $r_*$ in terms of which the
horizon is at $r_*=-\infty$ and $i_0$ is at $r_*=\infty$.  The question
is therefore, given a perturbation incoming from $r_*=\infty$, how much
energy makes it to $r_*=-\infty$? Scattering in the Zerilli potential
has been well studied, and one knows that for a generic incoming
perturbation from $r_*=\infty$ one gets a damped oscillation known as
quasinormal ringing at the horizon ($r_*=-\infty$),
\begin{equation}
\psi_{\rm Horizon} = A \exp(-\omega_i t) \sin(\omega_r t) \end{equation}
where the quasinormal frequencies scale as inverse powers of the
background spacetime mass, \begin{equation} \omega_{r,i} = {{\rm
constant}\over M_1} \end{equation} and the coefficient $A$ depends on
particular details of the initial perturbation.  A study of how
different perturbations generate different values of $A$ is found in
Price and Sun \cite{PrSu}.

The total energy falling into the black hole is proportional to the integral
$\int dt \dot{\psi}_{\rm Horizon}^2$. The dependence of this
integral on $M_1$ is determined completely by the behavior of the quasinormal
frequencies and gives as a result,
\begin{equation}
\Delta M \sim M_1^{-1}
\end{equation}
which implies $\beta=1/2$.

Let me end this note with some remarks.

$\bullet$ The above explanation, although rough, provides a link between
universality and no-hair behavior in gravitational collapse.

$\bullet$ In spite of the coarse approximation, the exponent
comes out with a value close to the experimentally measured one.  It
seems that the fact that the value is the same as in the exact solutions
or the Strominger-Thorlacius \cite{StTh} model is a mere coincidence,
since none of the features of those examples has been incorporated in
this discussion.

$\bullet$ The above mechanism predicts the same value for the exponent for
the scalar field studied by Choptuik, the gravitational waves studied by
Abrahams and Evans and the fluid case studied by Coleman and Evans, since
it only depends on the relation of the quasinormal frequencies to the
background spacetime mass, which does not depend on the details of the
perturbations.

The outstanding question is: is there any way to refine the approximation
in order to give the correct exponent? The following remarks are in order.

$\bullet$ The dependence of quasinormal frequencies on the background
mass is a robust feature, that seems unlikely to change if one modifies
the scenario.  It is possible that a dynamical background could alter
the picture of quasinormal ringing at the horizon and therefore alter
the exponent.  This would bring into question up to what extent is the
link between universality and baldness a mere artifact of our rough
approximation.  If quasinormal ringing were still present in a dynamical
background, in order to predict the right exponent, the frequencies
would have to depend non-analytically on the background mass.  This
seems quite unlikely, even on dimensional grounds.  Traschen \cite{Tr}
has found some non-analytic behavior in perturbations of Reissner
Nordstrom.  Perhaps there is a connection.  It could also happen that in
the non-static case there is a residual dependence of the coefficient $A$
on $M_1$, which in the static case does not appear.  This could allow
non-analytic behavior of the energy without a counter-intuitively
non-analytic behavior of the quasinormal frequencies.

$\bullet$ Insight into perturbations of dynamical backgrounds could be
gained by studying perturbations of the Vaidya metric. This would provide
a dynamical background and could help gain intuition on some of the issues
raised in the previous point.

$\bullet$ Finally, in the case of the collapse of radiative fluid,
studied by Coleman and Evans, the exact critical solution is known
analytically.  Studying perturbations of it should certainly allow
insight into the critical exponent.  Evans is currently pursuing this
point, with a rather different approach than the one presented here.

I wish to to thank Doug Eardley, Charles Evans, David Garfinkle,
Carsten Gundlach and Richard Price for useful comments and
discussions.  This work was supported by funds of the Office of
Minority Faculty Development at Penn State University.  I wish to
thank the Aspen Center for Physics and the Max Planck Institut f\"ur
Astrophysik (Munich) for hospitality during the preparation of this
manuscript.

\end{document}